\documentclass[a4paper,fleqn,usenatbib]{mnras}

\usepackage[T1]{fontenc}
\usepackage{ae,aecompl}

\usepackage{graphicx}	
\usepackage{amsmath}	
\usepackage{amssymb}	

\title[Properties of magneto-dipole lines]{Properties of magneto-dipole X-ray lines in different radiation models}

\author[G. S. Bisnovatyi-Kogan, Ya. S. Lyakhova]{
G. S. Bisnovatyi-Kogan,$^{1,2}$\thanks{E-mail: gkogan@iki.rssi.ru (GSBK)}
Ya. S. Lyakhova,$^{2,1}$
\\
$^{1}$Space Research Institute of Russian Academy of of Sciences, Profsoyuznaya 84/32, Moscow, 117997, Russia\\
$^{2}$National Research Nuclear University MEPhI (Moscow Engineering Physics Institute),\\ Kashirskoe Shosse 31, Moscow, 115409, Russia
}

\date{Accepted XXX. Received YYY; in original form ZZZ}

\pubyear{2015}

\begin{document}
\label{firstpage}
\pagerange{\pageref{firstpage}--\pageref{lastpage}}
\maketitle

\begin{abstract}
We compare polarization properties of the cyclotron, and relativistic dipole radiation of electrons moving in the magnetic field on a helix with ultra-relativistic longitudinal and non-relativistic transverse velocity components. The applicability of these models in the case of accretion onto a neutron star is discussed. The test, based on polarization observations is suggested, to distinguish between the cyclotron, and relativistic dipole origin of features, observed in X-ray spectra of some X-ray sources, among which the Her X-1 is the most famous.
\end{abstract}




\section{Introduction}

X-ray pulsar Hercules X-1  discovered in 1971 by the Uhuru satellite
is one of the best studied X-ray source. Her X-1 is the first source in which X-ray spectrum the line feature
in the 39-58 KeV energy range was observed, which could not be identified with any chemical element, and was
suggested to be a cyclotron line \cite{Trumper}. This feature was observed later in
\cite{Tueller, Voges, Ubertini, Gruber}. When
this feature is interpreted as a cyclotron line, the magnetic field strength
may be calculated from the non-relativistic formula

\begin{equation}
\label{bfield}
B=\frac{m_e c\omega}{e},
\end{equation}
where $\omega$ is the cycle frequency of the electrons, identified with the frequency of the observed X-ray feature,
$m_e$ is the mass of the electron, $c$ is the light speed. In this case
the magnetic field strength should be of the order of $(3 - 5)\times 10^{12}$ Gs. But as large as this value
comes into conflict with some theoretical reasonings among which the most important are consideration of the interrelation
between radio and X-ray pulsars \cite{BKKomberg}, and simulation of the pulse variability during the 35-days
cycle in observations from
the satellites ASTRON  \cite{Sheffer}, Ginga and RXTE \cite{Scott, Deeter}.
Obscuration of X-ray beams during the 35
day cycle is often used to explain the periodic
X-ray high-low state transitions of Her X-1 during the accretion disk precession. If the obscuring
material is the inner edge of the accretion disk, then the inner disk must be tilted out of the
binary plane and be precessing to produce periodically
varying obscuration. In such a situation, occultation of the
neutron star would occur twice in each precession cycle,
leading to the decline in flux, and termination of the main and
short high states. This scheme was also extended \cite{Sheffer} to
explain pulse profile evolution with a reflection of the light on the off state by the inner edge of the accretion disk.
The value of the dipole magnetic field of the neutron star, determining the radius of the inner edge, coinciding with the
radius of the Alfven surface, was estimated  in this model as $10^{10} - 10^{11}$~Gs.

Let us stress, that in this model the region where the non-collision shock wave is formed is situated at the upper side of the
accretion column, so it is separated in space from the region where the main X-ray flux is formed. Therefore, in connection
with this model, there is no need to make any principal changing in the standard model. Only  the structure of the accretion column could be
modified adjusting to the lower value of the magnetic field.

The most reliable estimation of the magnetic field $B_S$ in neutron stars (dipole component) is obtained for
radiopulsars by measurements of a growth of their rotation period at magneto-dipole losses. For single radiopulsars,
forming a large group of about 2000 objects,
this field is varies \citet{lorimer05} around $10^{12}$ G. In addition to the main body of the objects on the diagram
$(P,\dot P)$ there is a smaller group of radiopulsars \citet{lorimer05} with more rapid rotation and
lower magnetic fields $10^8\,-\,5\cdot 10^{10}$. About 200 of these pulsars are called "recycled pulsars", which
passed the stage of accretion in close binaries, when they gain a rapid rotational speed, and decrease their magnetic field
\cite{BKKomberg}. Majority of recycled pulsars had low mass companions, remaining as a white dwarf in the binary system, and lower
magnetic fields $10^8\,-\,10^9$ Gs, and few tens of pulsars are in the binary with another neutron star, and magnetic field up to
$\sim 5\cdot 10^10$ Gs. The optical companion of the Her X1 is a star with mass $\sim 2\, M_\odot$, which ends its evolution as
a white dwarf. Therefore, there is not surprising for the neutron star in this system to have a magnetic field in the range
$10^{10} - 10^{11}$~Gs.

To solve the problem of discrepancy between this estimation and the value following from the cyclotron
interpretation (\ref{bfield}), it was suggested in \cite{Baushev}, that the observed feature could be explained by the
relativistic dipole radiation of electrons having strongly anisotropic distribution function, with
ultra-relativistic motion along the
magnetic field lines, and non-relativistic motion across it. Such distribution function is formed when the
accretion flow into the
magnetic pole of the neutron star is stopped in a non-collisional shock wave \cite{BKF69}, and a rapid
loss of transversal energy
in the strong magnetic field leads to strongly  anisotropic momentum distribution \cite{BK73}.

It is not possible for the moment to make a definite choice between these two models. There are
another models explaining the change of X-ray beam during 35 day period without obscuration of the
beam by the inner edge of the accretion disk (Postnov et al., 2013; Staubert et al., 2013). Therefore
only observational criteria permit to make a choice between the models.

In this paper we consider the problem of the observational choice between the above mentioned models
by measuring the polarization of the
radiation in this X-ray feature. The relativistic dipole and cyclotron radiation have different
polarization properties, so such measurements
could solve this long-standing problem. Such experiments could be performed on the Japanese satellite
\textit{Astro-H} which launch is planned for 2015, \cite{AstroH}. For description of different ways of
X-ray polarization measurements
see \cite{polar2,xpolar}, and references therein.

\section{Polarization and emissivity of the cyclotron radiation}
\label{sec:Cycl}
\begin{figure}
\centering
\begin{minipage}{0.6\linewidth}
\center{\includegraphics[width=1\linewidth]{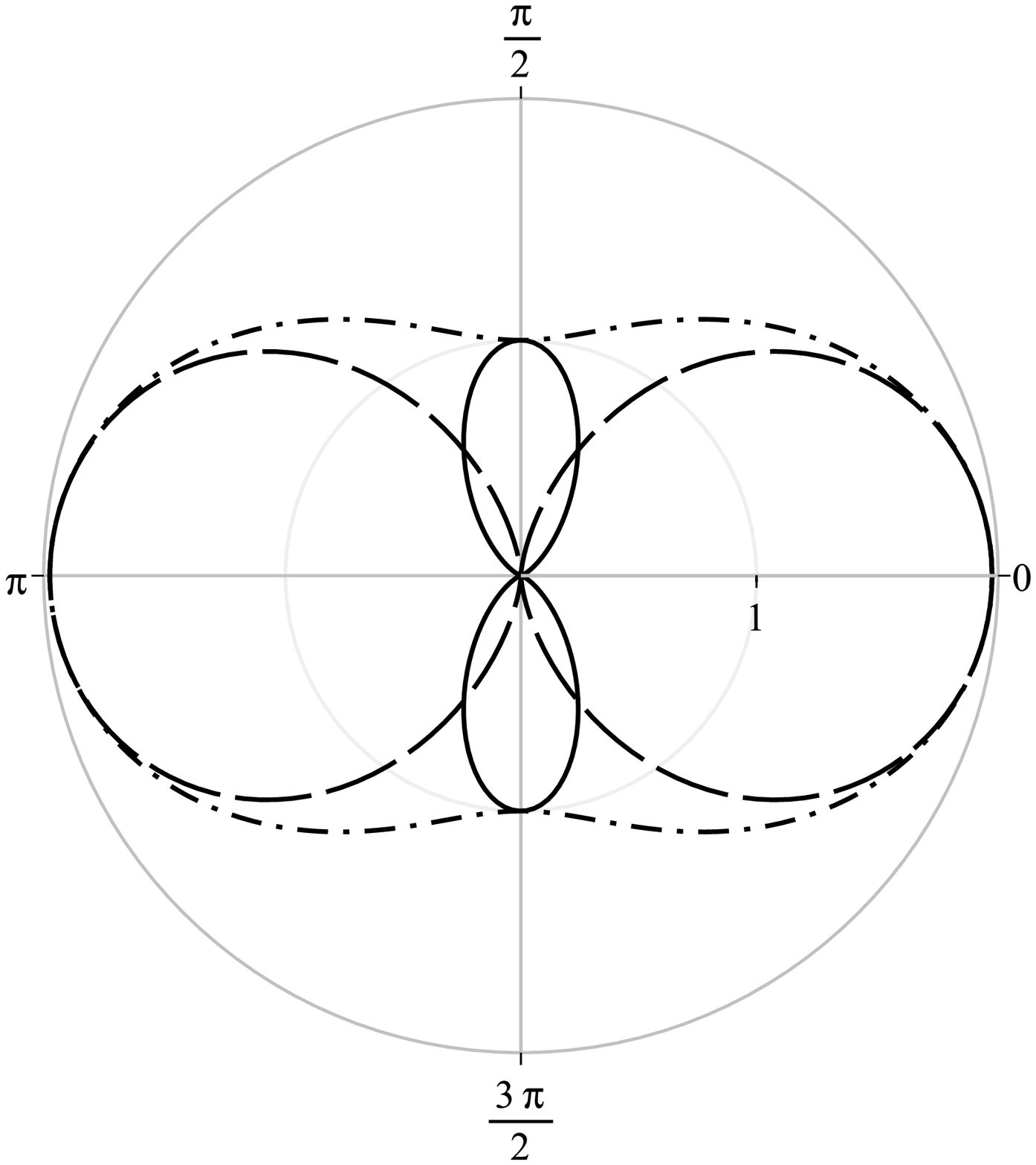}} \bigskip \\
\end{minipage}
\vfill
\begin{minipage}[h!]{0.6\linewidth}
\center{\includegraphics[width=1\linewidth]{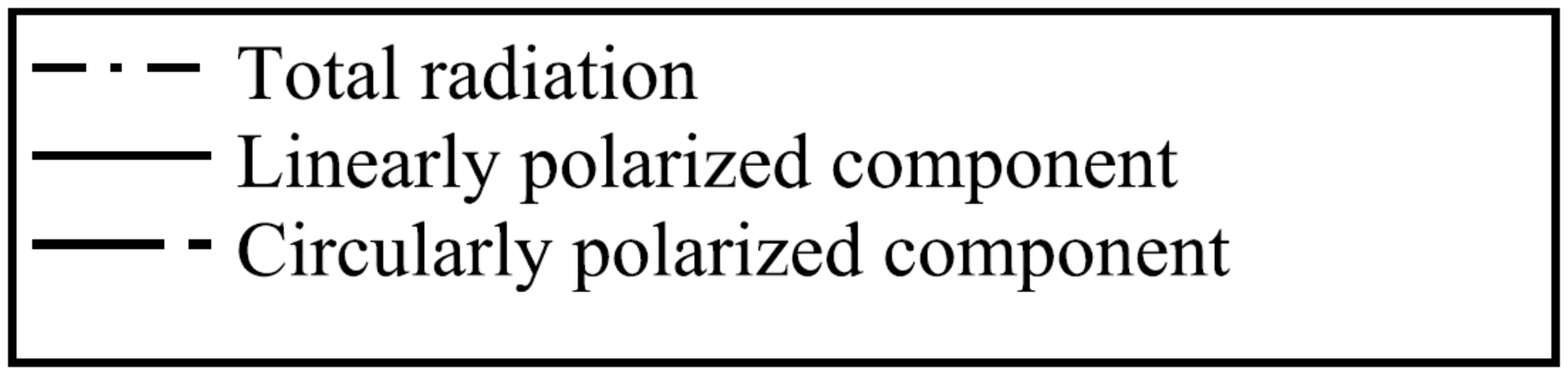}}
\\1.a)
\end{minipage}
\vfill
\begin{minipage}{0.6\linewidth}
\center{\includegraphics[width=1\linewidth]{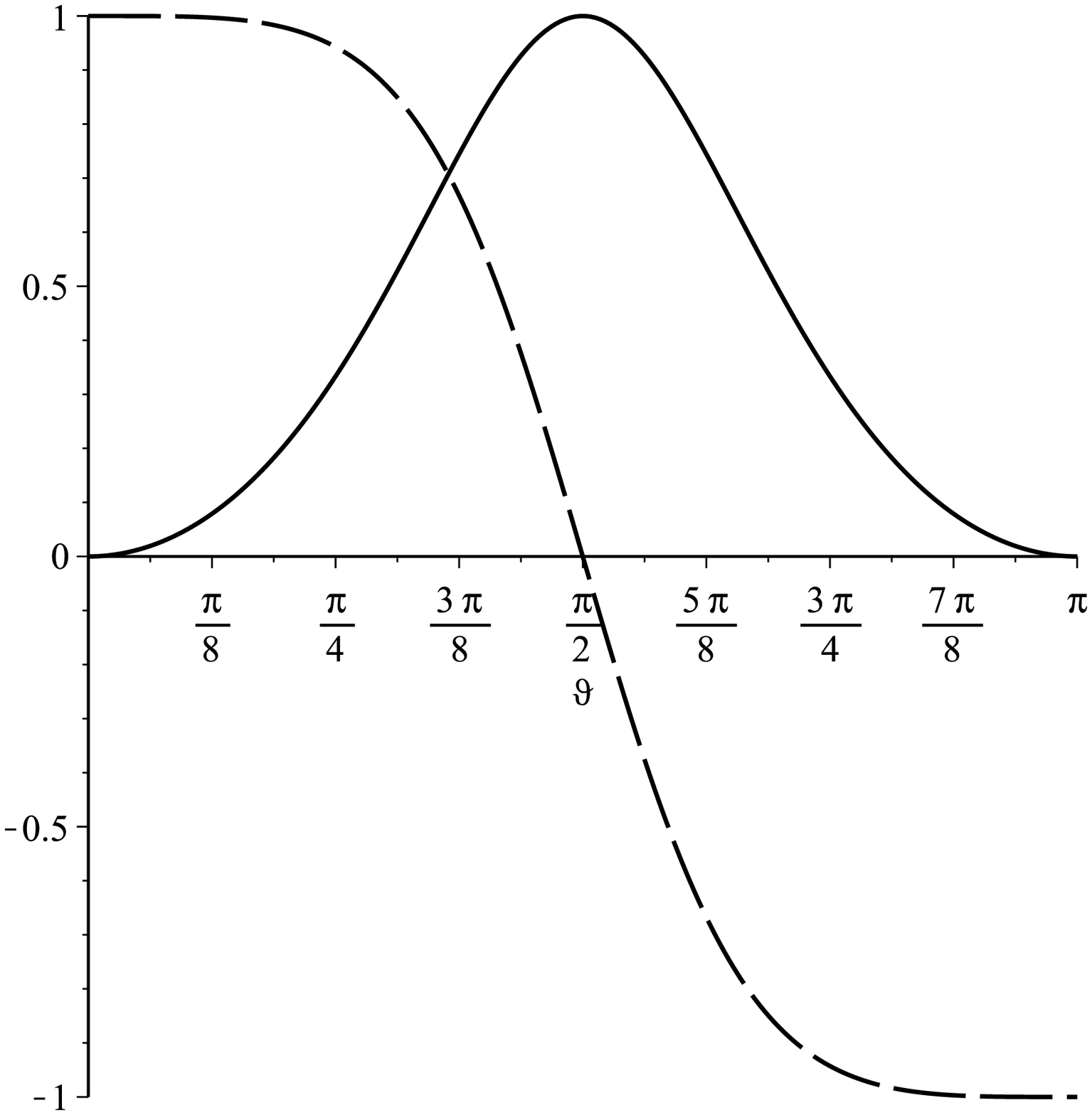}} \bigskip\\
\end{minipage}
\vfill
\begin{minipage}[h!]{0.6\linewidth}
\center{\includegraphics[width=1\linewidth]{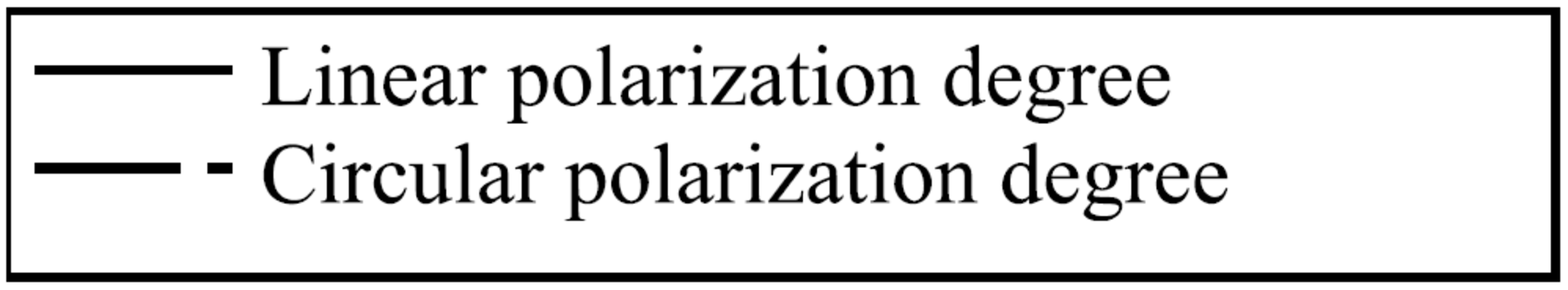}}
\\1.b)
\end{minipage}
\bigskip
\caption{Angular distribution of CR polarization components (a) and angular dependence of the linear and circular polarization degrees (b). Arbitrary units are used.}
\label{fig:CR_angle}
\end{figure}

The cyclotron radiation is produced during a motion of non-relativistic electrons across a magnetic field direction.
It is radiated in the form of the line with the energy $\hbar \omega_B$, with the cyclotron frequency

\begin{equation}
\label{eq:freq0}
\omega_{B} = \frac{eB}{m_e c}, \qquad \nu_0=\frac{\omega_B}{2\pi}.
\end{equation}
The electron is moving along the Larmor circle with the radius

\begin{equation}
\label{eq: larmor}
 R_{L} = \frac{m \upsilon_{\perp,0}}{eB},
\end{equation}
where the electron velocity $\upsilon_{\perp,0}$ is the component situated
in the plane perpendicular to the direction of the magnetic field in the frame connected with the Larmor circle.
The electron is radiating also on the harmonic frequency $\omega_{nB}=n\omega_B$. At $\upsilon_{\perp,0} \ll c$ the strength
of the harmonic
lines is rapidly decreasing with the number $n$. If also the total $|\upsilon| \ll c$, the change of cyclotron
frequency due to Doppler shifting may be neglected,
and only the gravitational redshift in the gravitational field of the neutron star
(not present in (\ref{bfield})) should be taken into account
for the magnetic field evaluation.
Taking into account only the radiation on the first harmonic of the cyclotron frequency, we have it's differential
angular emissivity $W_0(\vartheta)$ as \cite{trub}

\begin{equation}
\label{eq4}
\begin{split}
W_{0} = \frac{e^{2}\omega^{2}_B \upsilon_{\perp 0}^{2}}{8\pi c^{3}} \left (1 + \cos^{2}{\vartheta_0} \right )\delta(\omega-\omega_B) {\rm \frac{erg}{s\cdot sterad\cdot Hz}},
\\
\upsilon_{\perp 0}\ll c,
\end{split}
\end{equation}
and the total emissivity, after integration over the angle and frequency, is:

\begin{equation}
\label{eq5}
W_{tot} = \frac{2e^{2}\omega^{2}_B \upsilon_{\perp 0}^{2}}{3 c^{3}}\,  {\rm \frac{erg}{s}}.
\end{equation}
Expressions for the degrees of linear and circular polarization, respectively,  are written as \cite{Epstein}:

\begin{equation}
\rho_{l 0} = \frac{1 - \cos^{2}{\vartheta_0}}{1 + \cos^{2}{\vartheta_0}},
\label{lin}
\end{equation}

\begin{equation}
\rho_{c 0} = \frac{2\cos{\vartheta_0}}{1 + \cos^{2}{\vartheta_0}}, \qquad \rho_{l {0}}^2 + \rho_{c {0}}^2=1.
\label{circ}
\end{equation}
The cyclotron radiation of a single electron is totally polarized, inducing the last equality.
 The cyclotron radiation along the
direction of the magnetic field is fully circularly polarized and in the plane perpendicular to the magnetic field it's fully linearly polarized.
We shall use the subscript "0" for the frame, connected with the plane of the Larmor circle where $\upsilon_{{\parallel} 0}=0$.
Angular distribution of the emissivity: full $W_0$ from (\ref{eq4}), polarized linearly $W_{0l}$, and circularly $W_{0r}$ of a cyclotron radiation
are presented in Fig. \ref{fig:CR_angle}.
The linear and circular emissivities are determined as

\begin{equation}
W_{0l}(\vartheta_0)=W_0 \rho_{l 0}^2, \qquad W_{0r}(\vartheta_0)=W_0 \rho_{r 0}^2,
\label{emlr}
\end{equation}
where $\rho_{l 0}$ and $\rho_{r 0}$ are given in (\ref{lin}) and (\ref{circ}), respectively.
The angle $\vartheta_0=0$ corresponds to the direction of the magnetic field.


\section{Polarization and emissivity of the relativistic dipole radiation}
\label{sec:MDM}

Let's consider an electron in the magnetic field, with the following values of  the velocity components in the
laboratory frame

$$
\upsilon_{\parallel} \simeq c, \qquad
\gamma_{\parallel} = \frac{1}{\sqrt{1 - \frac{\upsilon_{\parallel}^{2}}{c^{2}}}} \gg 1,
$$
\begin{equation}
\label{rdr}
\upsilon_{\perp} \ll c
\sqrt{1 - \frac{\upsilon_{\parallel}^{2}}{c^{2}}}
=\frac{c}{\gamma_{\parallel}}.
\end{equation}

\begin{figure}
\centering
\includegraphics[scale=0.3]{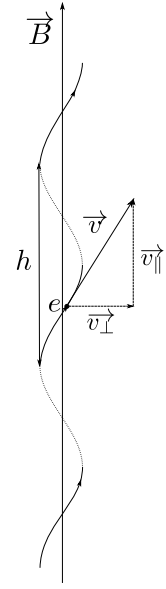}
\caption{The RDR trajectory: electron moves on helix along the magnetic field; $h$ is the helix step.}
\label{fig:MDRtr}
\end{figure}

\noindent The trajectory of the electron is helical, with the helix step  significantly larger than it's radius
(see Fig. \ref{fig:MDRtr}).

\begin{figure}
\centering
\begin{minipage}[h!]{0.6\linewidth}
\center{\includegraphics[width=1\linewidth]{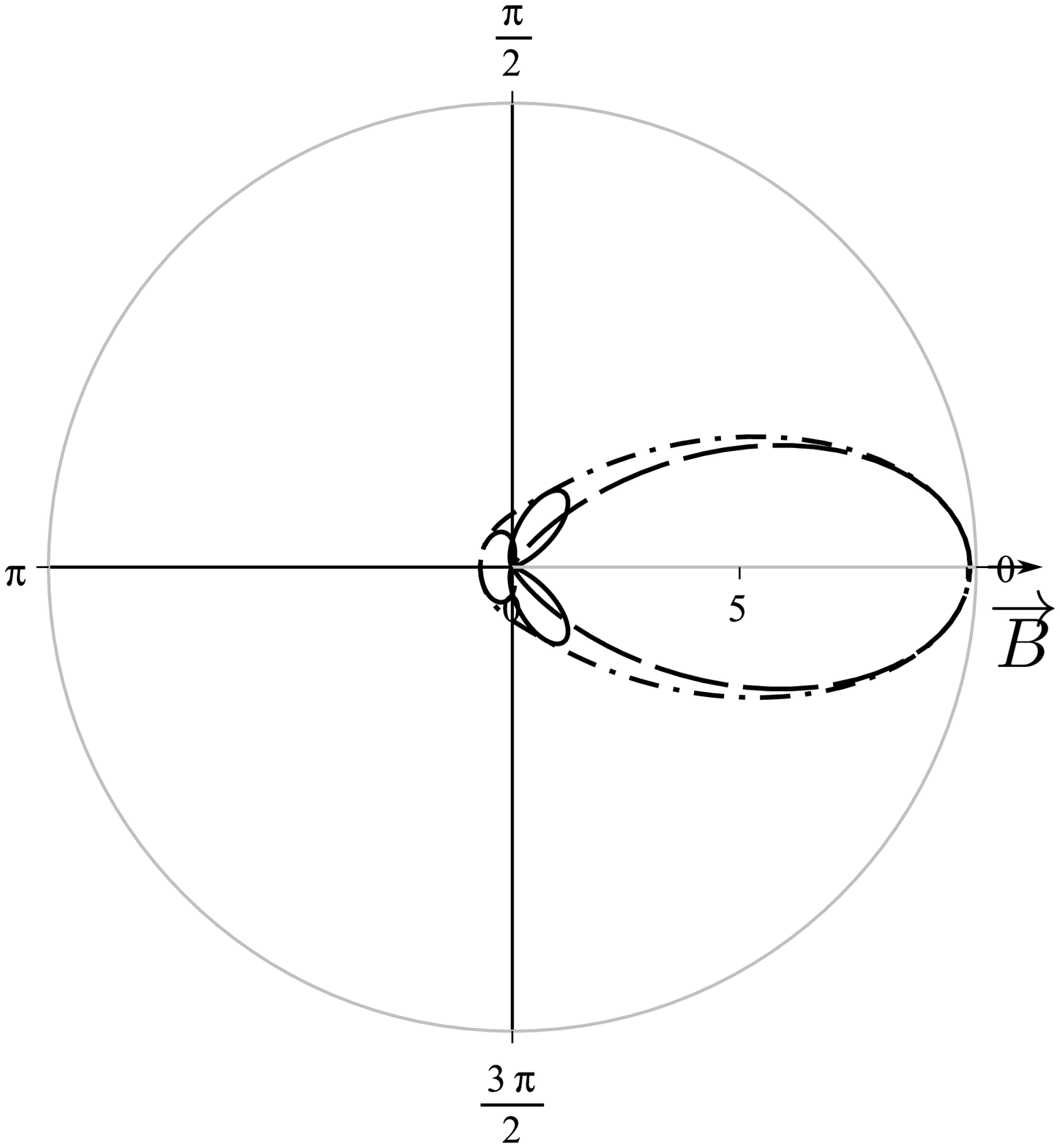}} \bigskip\bigskip \\3.1
\end{minipage}
\vfill
\begin{minipage}[h!]{0.6\linewidth}
\center{\includegraphics[width=1\linewidth]{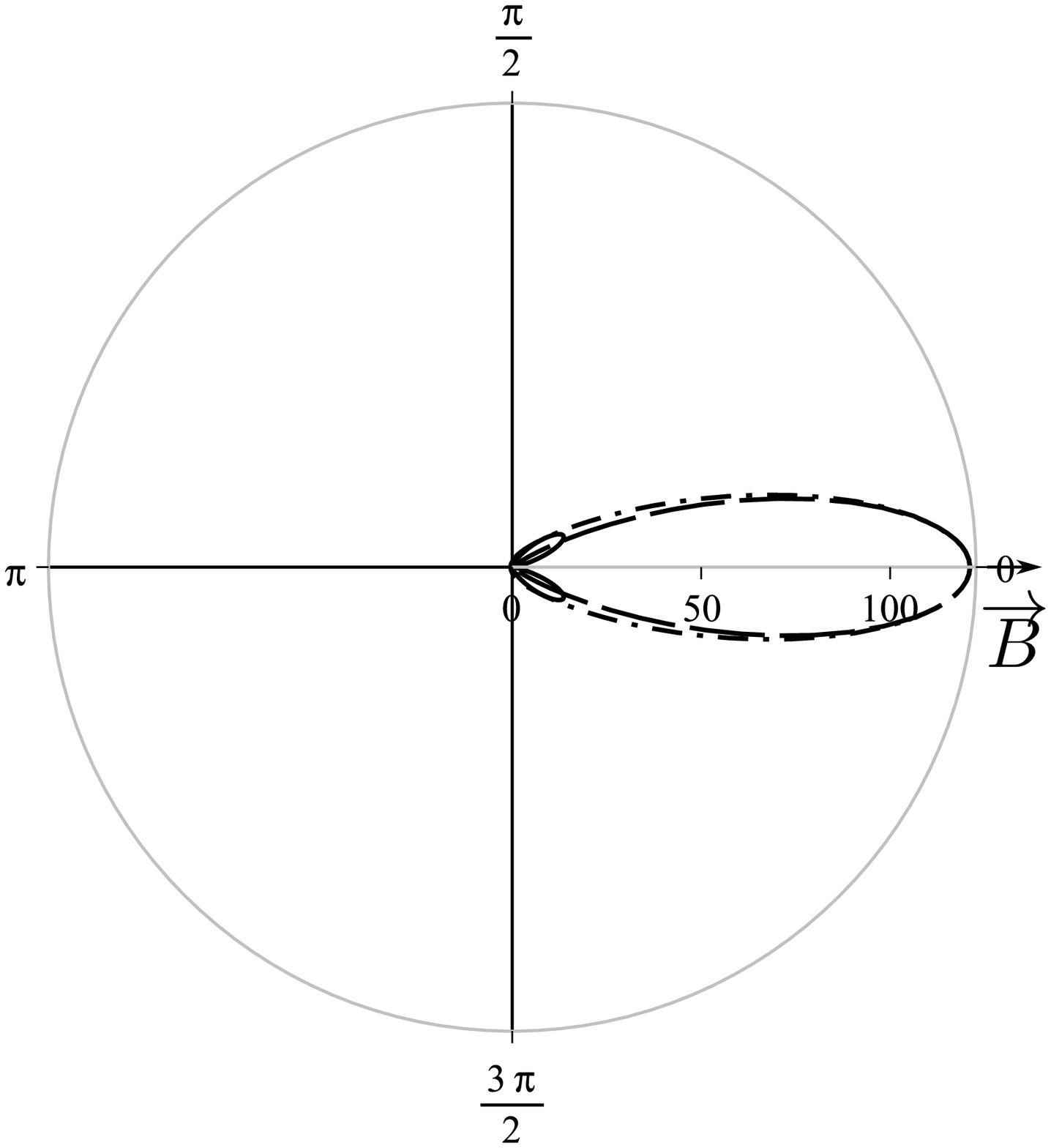}} \bigskip\bigskip \\3.2
\end{minipage}
\vfill
\begin{minipage}[h!]{0.6\linewidth}
\center{\includegraphics[width=1\linewidth]{Caption.eps}} \bigskip\bigskip\\
\end{minipage}
\bigskip
\caption{Angular distribution of RDR polarization components for different values of Lorentz
parameters: 3.1) $\gamma_{\parallel} = 1.1$; 3.2) $\gamma_{\parallel} = 1.5$. Arbitrary units are used.}
\label{fig:MDRangle1}
\end{figure}

\begin{figure}
\centering
\begin{minipage}[h!]{0.6\linewidth}
\center{\includegraphics[width=1\linewidth]{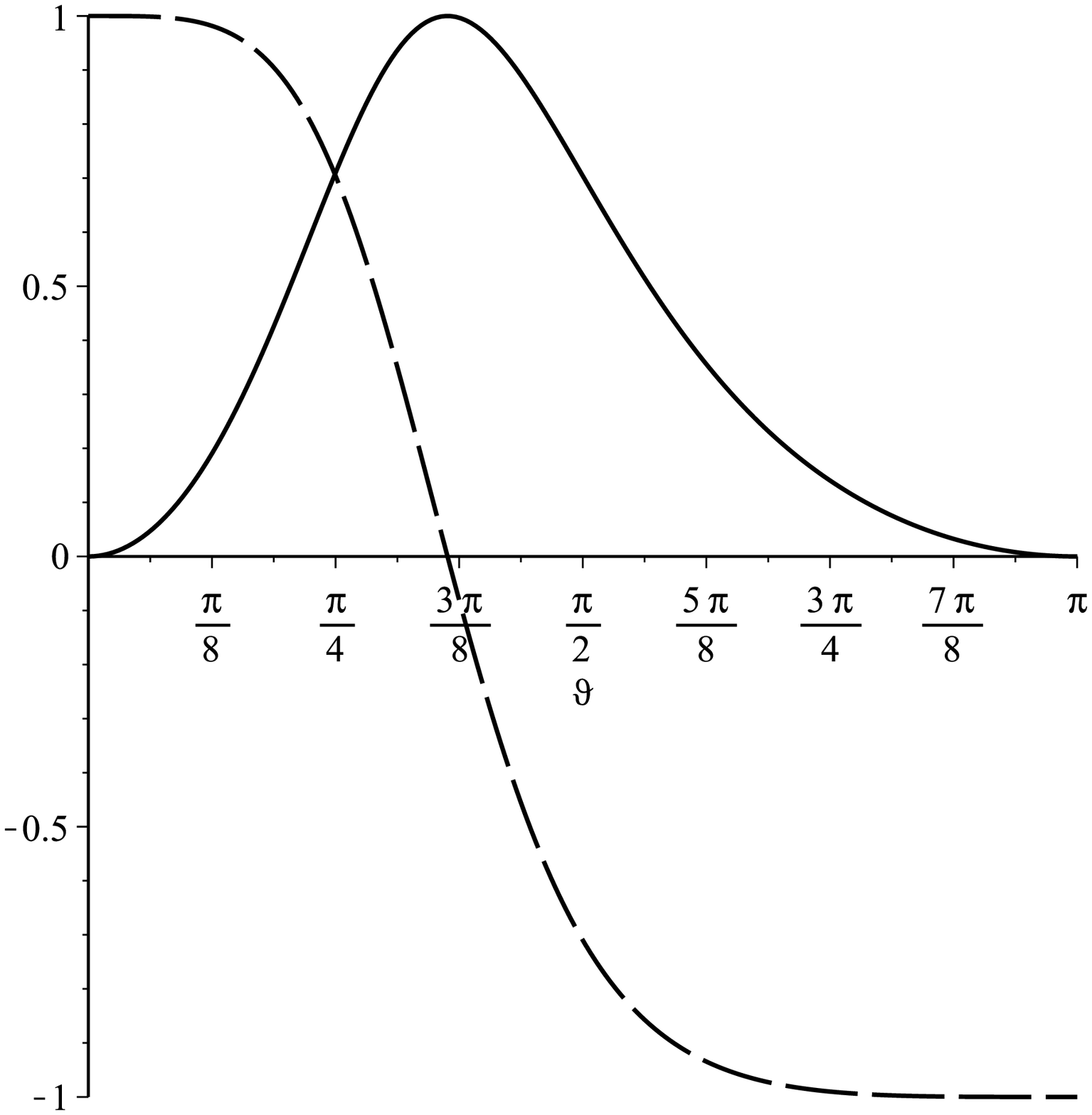}} \bigskip\bigskip\\4.1
\end{minipage}
\vfill
\begin{minipage}[h!]{0.6\linewidth}
\center{\includegraphics[width=1\linewidth]{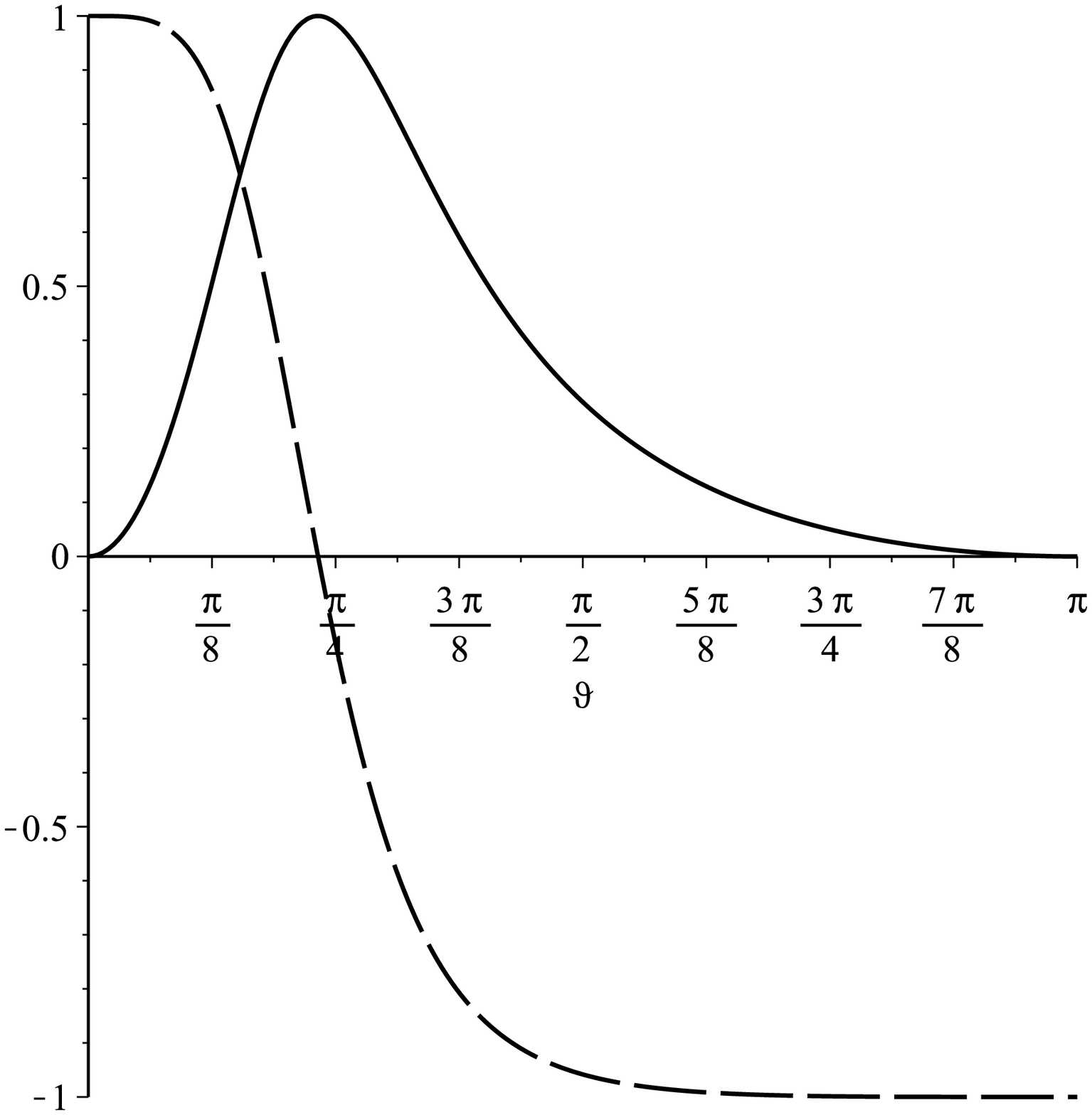}} \bigskip\bigskip \\4.2
\end{minipage}
\vfill
\begin{minipage}[h!]{0.6\linewidth}
\center{\includegraphics[width=1\linewidth]{Caption2.eps}}\bigskip\bigskip \\
\end{minipage}
\bigskip
\caption{Angular dependence of the linear and circular polarization degrees for different values of Lorentz parameters: 4.1) $\gamma_{\parallel} = 1.1$; 4.2) $\gamma_{\parallel} = 1.5$. Arbitrary units are used.}
\label{fig:MDRangle2}
\end{figure}

\begin{figure}
\centering
\begin{minipage}[h!]{0.6\linewidth}
\center{\includegraphics[width=1\linewidth]{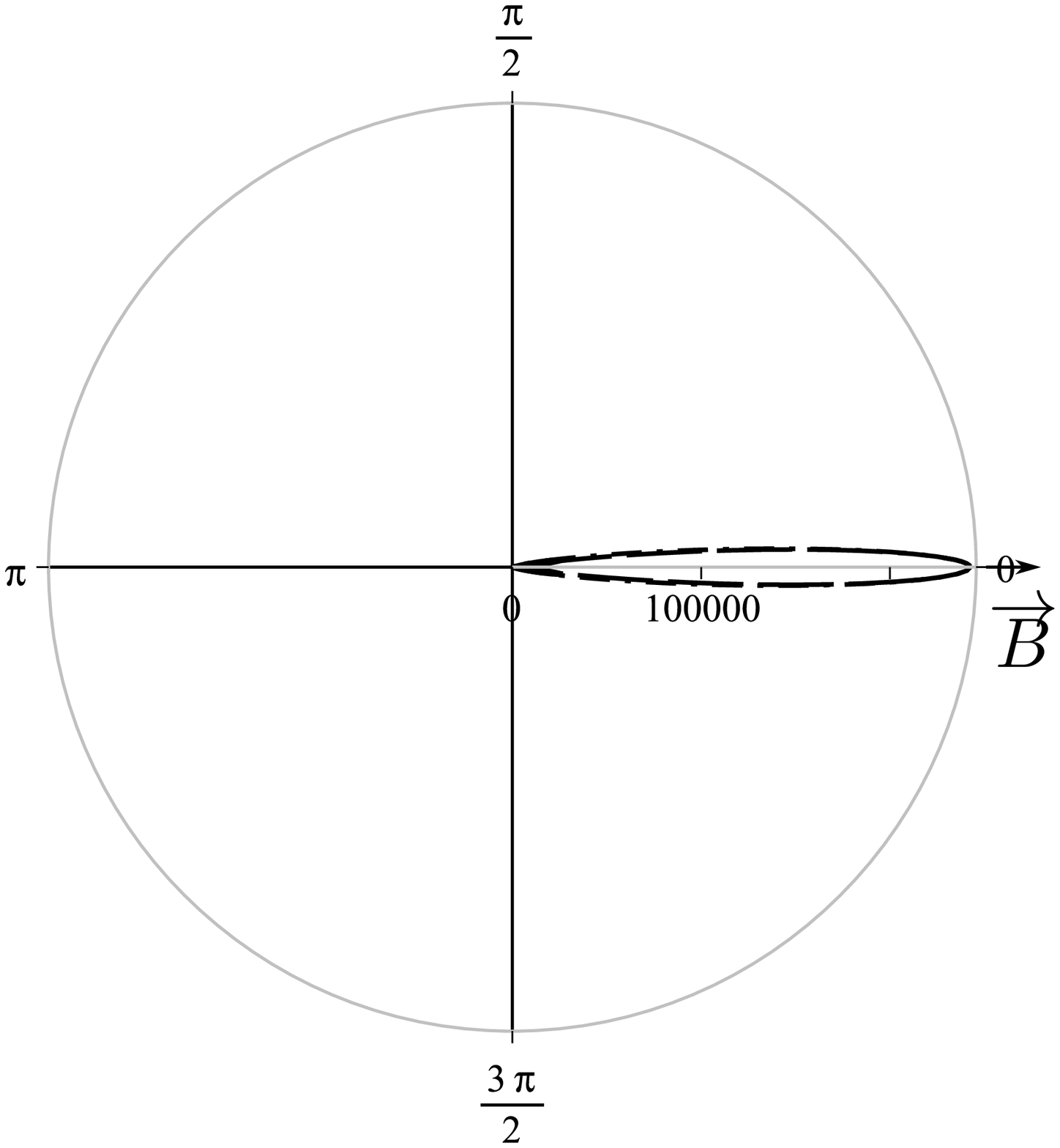}} \bigskip\bigskip \\5.1
\end{minipage}
\vfill
\begin{minipage}[h!]{0.6\linewidth}
\center{\includegraphics[width=1\linewidth]{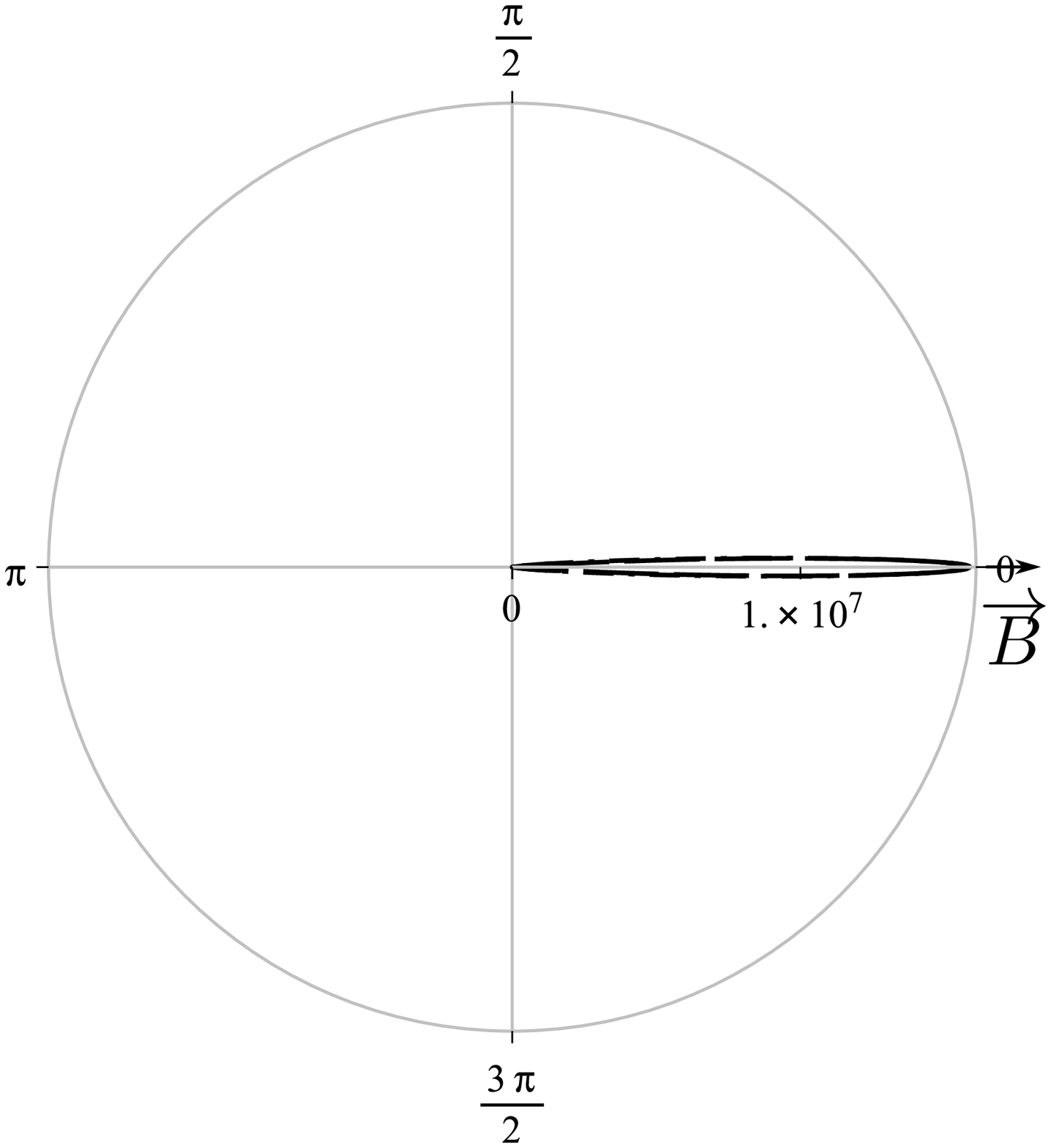}} \bigskip\bigskip \\5.2
\end{minipage}
\vfill
\begin{minipage}[h!]{0.6\linewidth}
\center{\includegraphics[width=1\linewidth]{Caption.eps}} \bigskip\bigskip\\
\end{minipage}
\bigskip
\caption{Angular distribution of RDR polarization components for different values of Lorentz
parameters: 5.1) $\gamma_{\parallel} = 3.0$; 5.2) $\gamma_{\parallel} = 10.0$. Arbitrary units are used.}
\label{fig:MDRangle3}
\end{figure}

\begin{figure}
\centering
\begin{minipage}[h!]{0.6\linewidth}
\center{\includegraphics[width=1\linewidth]{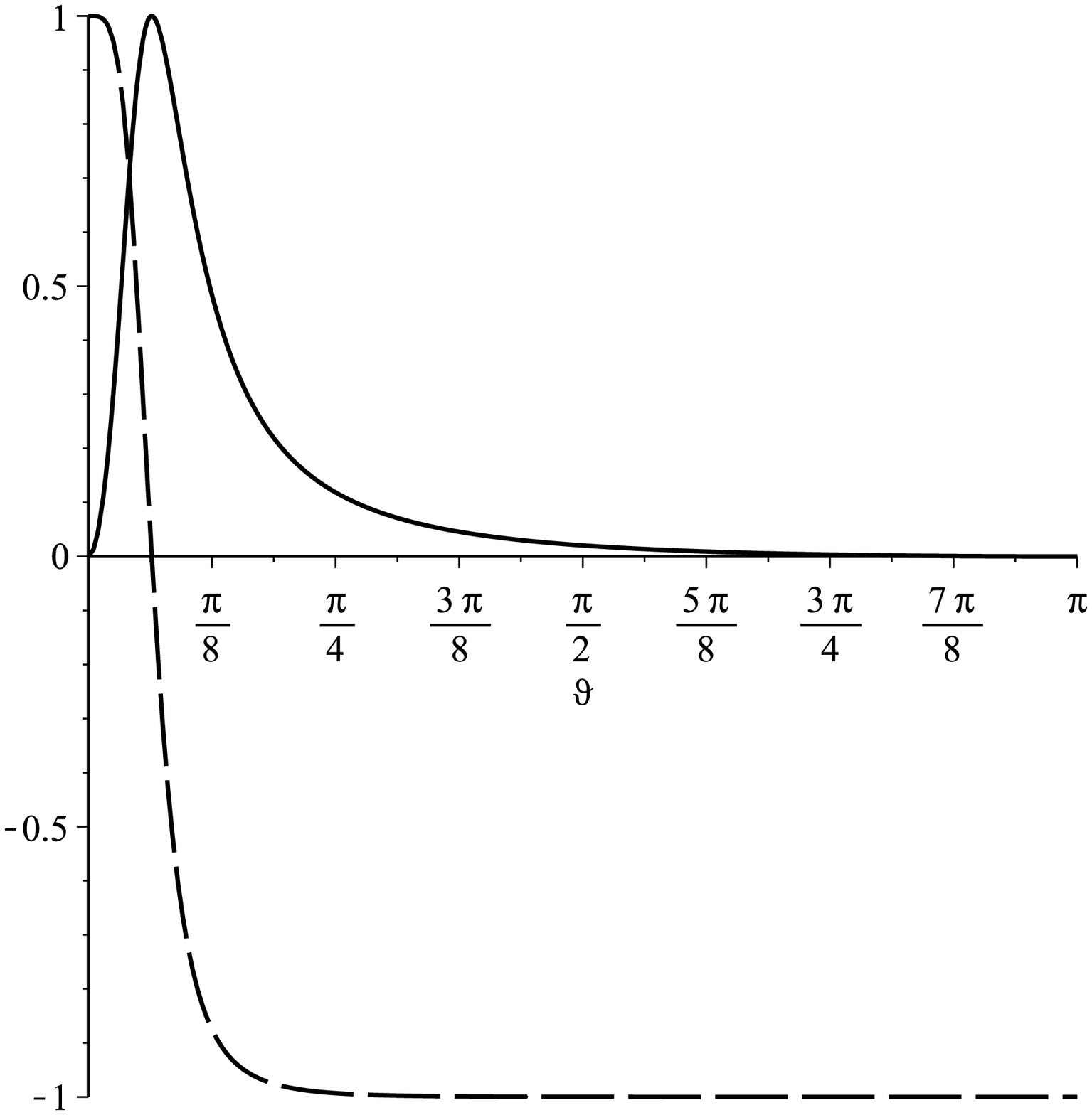}} \bigskip\bigskip\\6.1
\end{minipage}
\vfill
\begin{minipage}[h!]{0.6\linewidth}
\center{\includegraphics[width=1\linewidth]{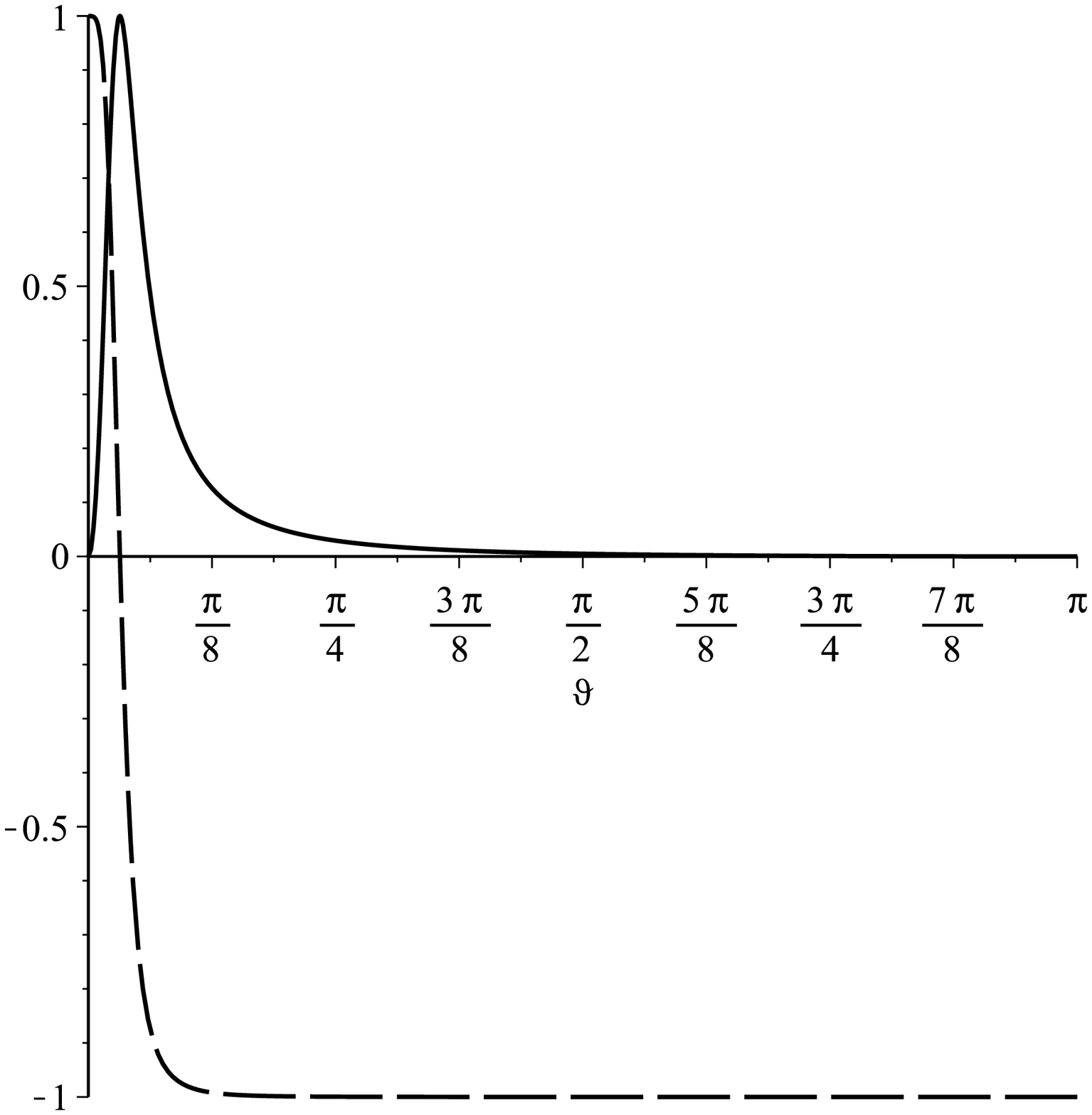}} \bigskip\bigskip \\6.2
\end{minipage}
\vfill
\begin{minipage}[h!]{0.6\linewidth}
\center{\includegraphics[width=1\linewidth]{Caption2.eps}}\bigskip\bigskip \\
\end{minipage}
\bigskip
\caption{Angular dependence of the linear and circular polarization degrees for different values of Lorentz parameters: 6.1) $\gamma_{\parallel} = 3.0$; 6.2) $\gamma_{\parallel} = 10.0$. Arbitrary units are used.}
\label{fig:MDRangle4}
\end{figure}

The  radiation provided by such system is called \cite{zhelezn} Relativistic Dipole (RDR). The properties of RDR
have been considered in detail in \cite{Epstein}.
The calculations of the angular distributions of RDR emissivity power, and both types of polarization
in the laboratory frame, where the electron is moving along the magnetic field to the observer with the
velocity $\upsilon_\parallel$, may be calculated by
making Lorentz transformation  in (\ref{eq4}),(\ref{lin}),(\ref{circ}). The angle $\vartheta_0$ and velocity
$\upsilon_{\perp 0}$ in the Larmor circle frame
are connected with the angle $\vartheta$ and velocity $\upsilon_{\perp}$
in the laboratory frame as ($\beta_\parallel=\upsilon_\parallel/c)$

\begin{equation}
\label{w}
\begin{split}
\sin\vartheta_0=\frac{\sin\vartheta \sqrt{1-\beta_\parallel^2}}{1 - \beta_{\parallel}\cos{\vartheta}},\qquad
\cos{\vartheta_0}=\frac{{\cos{\vartheta}-\beta_\parallel}}{1 - \beta_{\parallel}\cos{\vartheta}},\\ \upsilon_{\perp 0}=\gamma_\parallel \upsilon_\perp.
\end{split}
\end{equation}
Expressions for the linear and circular polarization degrees in the laboratory frame are obtained from (\ref{lin}),(\ref{circ}),
with account of (\ref{w}), as

\begin{equation}
\label{eq8}
\rho_{l} = \frac{1-\left(\frac{\cos{\vartheta}-\beta_\parallel}{1-\beta_{\parallel}\cos{\vartheta}}\right)^{2}}
{1+\left(\frac{\cos{\vartheta}-\beta_\parallel}{1-\beta_{\parallel}\cos{\vartheta}}\right)^{2}},
\qquad
\rho_{c} = \frac{2\frac{\cos{\vartheta}-\beta_\parallel}{1-\beta_{\parallel}\cos{\vartheta}}}
{1+\left(\frac{\cos{\vartheta}-\beta_\parallel}{1-\beta_{\parallel}\cos{\vartheta}}\right)^{2}}.
\end{equation}
It follows from (\ref{eq4}), (\ref{w}) that RDR radiation is emitted in a small angle ($\vartheta\lesssim 1/\gamma_\parallel$), along
the magnetic field direction. We have, by definition, $\beta_{\parallel}^2=1-(1/\gamma_{\parallel}^2)$.
For small $\vartheta$, and large $\gamma_\parallel$ we have the following expansions

$$
\beta_\parallel \approx 1-\frac{1}{2\gamma_\parallel^2}-\frac{1}{8\gamma_\parallel^4},\quad
$$
\begin{equation}
\cos\vartheta\approx 1-\frac{\vartheta^2}{2}+\frac{\vartheta^4}{24},\quad
\cos^2\vartheta\approx 1-\vartheta^2+\frac{\vartheta^4}{3},
\label{appr}
\end{equation}
It is convenient \cite{RTRP}, to introduce a variable

\begin{equation}
\label{psi}
\psi = \gamma_{\parallel}\vartheta.
\end{equation}
With account of (\ref{psi}) we obtain from (\ref{appr}) the expressions

\begin{equation}
1 - \beta_{\parallel}\cos{\vartheta} \approx \frac{1}{2\gamma
_{\parallel}^{2}}\left(1 + \psi^{2}\right),
\qquad
 \cos{\vartheta}-\beta_{\parallel} \approx \frac{1}{2\gamma_{\parallel}^{2}}\left(1 - \psi^{2}\right).
\label{psi1}
\end{equation}
The emission of RDR is monochromatic in the laboratory frame in any given direction, with the frequency and
polarization depending on the angle $\theta$.
The angular frequency distribution is obtained from the relations for Doppler effect \cite{LLfield}
(see also \cite{Epstein}), which,  with account of (\ref{psi1}), are written for  the
time interval $dt$ and the frequency $\omega$, as

\begin{equation}
\quad dt_0=dt\sqrt{1-\beta_{\parallel}^2},
\label{psi2}
\end{equation}
$$
\omega=\omega_B\frac{\sqrt{1-\beta_{\parallel}^2}}{1-\beta_{\parallel}\cos\vartheta},\quad
2\gamma_\parallel\omega_B \ge \omega \ge \frac{\omega_B}{2\gamma_\parallel} \quad {\rm at}\quad 0\le \vartheta \le\pi.
$$
Approximately we have
\begin{equation}
\omega\approx \frac{2\gamma_{\parallel}\omega_{B}}{1 + \psi^{2}}, \quad
\psi^2=\frac{2\gamma_\parallel\omega_B}{\omega}-1.
\label{psi2a}
\end{equation}
Angular dependencies of the polarization degrees in the laboratory frame from (\ref{eq8}), with account of (\ref{psi}),(\ref{psi2})
are written as  \cite{Epstein}:

\begin{equation}
\label{psi2b}
\rho_{l}(\omega) \approx 2\frac{\psi^{2}}{1 + \psi^{4}}=
\frac{\omega\left(2\gamma_{\parallel}\omega_{B} - \omega\right)}{\omega^{2} - 2\gamma_{\parallel}\omega\omega_{B} + 2\gamma_{\parallel}^{2}\omega_{B}^{2}},
\end{equation}
\begin{equation}
\label{psi2c}
\rho_{c}(\omega) \approx \frac{1 - \psi^{4}}{1 + \psi^{4}}=
\frac{2\gamma_{\parallel}\omega_{B}\left(\omega - \gamma_{\parallel}\omega_{B}\right)}{\omega^{2} - 2\gamma_{\parallel}\omega\omega_{B}
+ 2\gamma_{\parallel}^{2}\omega_{B}^{2}}.
\end{equation}
The relative graphics of angular dependencies are presented on Figs \ref{fig:MDRangle1} -- \ref{fig:MDRangle4}.

The differential power of the radiation in the unity of the solid angle $\Omega$, with $d\Omega=\sin\vartheta d\vartheta d\varphi$,
time $t$, and frequency is obtained from (\ref{eq4}), with account of (\ref{w}), (\ref{psi2}), and  relations

\begin{equation}
\label{psi3}
d\Omega_0=d\Omega\frac{d\cos\vartheta_0}{d\cos\vartheta}
=d\Omega\frac{1-\beta_{\parallel}^2}{(1-\beta_{\parallel}\cos\vartheta)^2},
\end{equation}
We have than from (\ref{eq4}), using (\ref{w}),(\ref{psi2}),
 the expression for the differential emissivity in the laboratory frame as (see \cite{Epstein})

\begin{equation}
\label{psi4}
\begin{split}
W=W_{0}\frac{(1-\beta_{\parallel}^2)^{3/2}}{(1-\beta_{\parallel}\cos\vartheta)^2}
=\frac{e^{2}\omega^{2}_B \upsilon_{\perp 0}^{2}}{8\pi c^{3}} \left [1 +
\frac{({\cos{\vartheta}-\beta_\parallel})^{2}}{(1 - \beta_{\parallel}\cos{\vartheta})^{2}} \right ]
\\
\times
\frac{(1-\beta_{\parallel}^2)^{3/2}}{(1-\beta_{\parallel}\cos\vartheta)^2}
\delta[\gamma_\parallel \omega(1-\beta_\parallel\cos\vartheta)-\omega_B].
\end{split}
\end{equation}
After transformation of $\delta$-function we have

\begin{equation}
\label{psi51}
W=\frac{e^{2}\omega^{2}_B \upsilon_{\perp 0}^{2}}{8\pi c^{3}\gamma_\parallel}
\left [1 +
\frac{({\cos{\vartheta}-\beta_\parallel})^{2}}{(1 - \beta_{\parallel}\cos{\vartheta})^{2}} \right ]
\end{equation}
$$
\times
\frac{(1-\beta_{\parallel}^2)^{3/2}}{(1-\beta_{\parallel}\cos\vartheta)^3}
\delta\left[ \omega-\frac{\omega_B}{\gamma_\parallel(1-\beta_\parallel\cos\vartheta)}\right].$$
Approximately we have

\begin{equation}
\label{psi4a}
W \approx\frac{2 e^{2}\omega^{2}_B \gamma_\parallel^2 \upsilon_{\perp 0}^{2}}{\pi c^{3}}\frac{1+\psi^4}{(1+\psi^2)^5} \delta\left( \omega-\frac{2\gamma_\parallel\omega_B}{1+\psi^2}\right).
\end{equation}
Using (\ref{psi2a}) we obtain in the laboratory frame \cite{Epstein})

\begin{equation}
\label{psi5}
W\approx\frac{e^{2} \upsilon_{\perp 0}^{2}\omega^3}{4\pi c^{3}\gamma_\parallel\omega_B}
\left(1-\frac{\omega}{\gamma_\parallel \omega_B} + \frac{\omega^2}{2\gamma_\parallel^2 \omega_B^2}\right)
\end{equation}
$$
\times\delta\left( \omega-\frac{2\gamma_\parallel\omega_B}{1+\psi^2}\right)
{\rm \frac{erg}{s\cdot sterad\cdot Hz}}.
$$
The spectral distribution of RDR is obtained after integration over $d\Omega$

\begin{equation}
\label{psi10}
W(\omega)=\int_{\Omega}W\,d\Omega\,\,{\rm \frac{erg}{s\cdot Hz}},
\end{equation}
and the total emissivity
\begin{equation}
W_{tot}=\int_0^{\omega_{max}}W(\omega)\,d\omega\,\,{\rm \frac{erg}{s}}.
\end{equation}
Taking into account (\ref{psi}),(\ref{psi2a}), we have  the relations for $\delta$-function as
\begin{equation}
\label{psi11}
2\pi\delta\left( \omega-\frac{2\gamma_\parallel\omega_B}{1+\psi^2}\right)\sin\theta d\theta d\omega \approx \pi \delta\left( \omega-\frac{2\gamma_\parallel\omega_B}{1+\psi^2}\right)d\theta^2 d\omega
\end{equation}
$$
=
\pi \frac{1+\psi^2}{\gamma_\parallel^2\omega}\delta\left( 1+\psi^2 -\frac{2\gamma_\parallel\omega_B}{\omega}\right)d(\psi^2+1) d\omega,
$$
After integration in (\ref{psi10}), with account of (\ref{psi11}), we obtain

\begin{equation}
\label{psi12}
W(\omega)=\frac{e^{2} \upsilon_{\perp 0}^{2}\omega^2}{4c^{3}\gamma_\parallel^3\omega_B}
\left(1-\frac{\omega}{\gamma_\parallel \omega_B} + \frac{\omega^2}{2\gamma_\parallel^2 \omega_B^2}\right)\times
\frac{2\gamma_\parallel\omega_B}{\omega}
\end{equation}
$$
= \frac{e^{2} \upsilon_{\perp 0}^{2}\omega}{2c^{3}\gamma_\parallel^2}
\left(1-\frac{\omega}{\gamma_\parallel \omega_B} + \frac{\omega^2}{2\gamma_\parallel^2 \omega_B^2}\right).
$$
The total emissivity, defining the rate of the energy loss of a particle, is obtained after integration of (\ref{psi12}), with  $\omega_{max}=2\gamma_\parallel\omega_B$ from (\ref{psi2}) as

\begin{equation}
\label{psi13}
W_{tot}=\int_0^{\omega_{max}}W(\omega)\,d\omega
=\frac{2e^{2} \upsilon_{\perp 0}^{2}\omega_B^2}{3c^{3}}\,\, {\rm \frac{erg}{s}},
\end{equation}
in accordance with (\ref{eq5}).

\section{Observational properties of relativistic dipole radiation}

We consider here, that the angular size of the aperture of the X-ray detector $\Delta \vartheta_X$ is much larger than the characteristic beam width of RDR
$\vartheta_{RDR}= 1/\gamma_\parallel$, so that
$\Delta \vartheta_X \gg 1/\gamma_\parallel$, $\gamma_\parallel\gg 1$. If the angular size of the hot spot on the neutron star
magnetic poles $\vartheta_{hs} \gg \vartheta_{RDR}$, than the registered X-ray RDR radiation is coming from the part of the hot spot,
and is lasted during the time $\tau_{RDR}\approx P\,\vartheta_{hs}/2\pi$. It is decreasing abruptly outside this time interval.
Note that in the case of ordinary cyclotron radiation its intensity is decreasing smoothly because of almost isotropic
radiation diagram $\sim (1 + \cos^{2}{\vartheta_0})$, according to (\ref{eq4}).

\subsection{RDR line widening}

Let us mention 3 mechanisms of the line widening in the magneto dipole radiation.

1) A variable magnetic field $B$ in the region of the line formation. This mechanism is working in the cyclotron radiation of
nonrelativistic electrons, emitting the line at frequency $\omega=\omega_B=\frac{eB}{m_e c}$, as well as in RDR where the
frequency of  radiation in the frame of the Larmor circle is changing similarly.

Two other mechanisms are characteristic only to the RDR.

2) The distribution of electrons over the parallel momentum $p_\parallel$  may be presented as \cite{Baushev}

\begin{equation}
\label{fppar}
f_e(p_\parallel)=\frac{n_e}{\sqrt{\pi\sigma}}\exp\left(-\frac{(p_\parallel-a)^2}{\sigma^2}\right).
\end{equation}
Such distribution of electrons leads to widening of the line as \cite{Baushev}

\begin{equation}
\label{ppar1}
\Delta\omega=\omega_B\frac{2\sigma}{m_e c}.
\end{equation}
For narrow momentum distribution of electrons $\sigma\ll a$ we have

\begin{equation}
\label{ppar2}
\Delta\omega \ll \omega_B\frac{2a}{m_e c}\approx 2\omega_B\gamma_\parallel.
\end{equation}
In this case the line width may be very narrow. If $\sigma< \frac{m_e c}{2}$, the width of the line will be

\begin{equation}
\label{ppar3}
\frac{\Delta\omega}{\omega}< \frac{1}{\gamma_\parallel}.
\end{equation}

\begin{figure}
\centering
\includegraphics[width=\columnwidth]{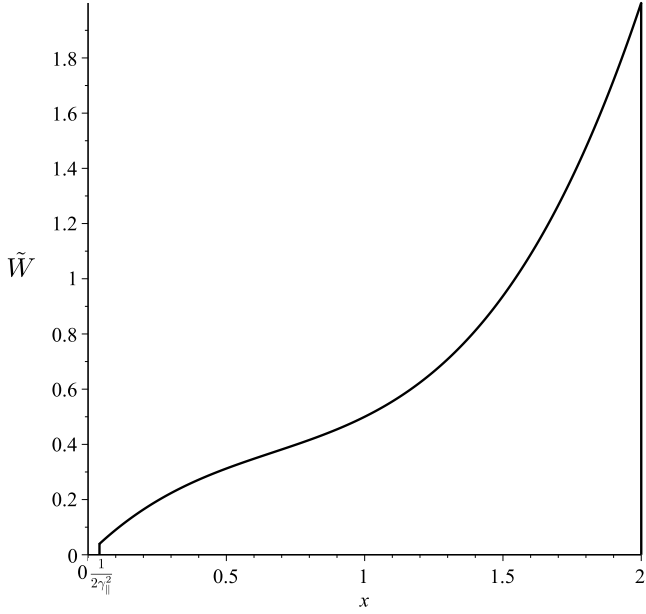}
\caption{The frequency distribution of energy in the RDR line.}
\label{fig:line}
\end{figure}

\begin{figure}
\centering
\includegraphics[width=\linewidth]{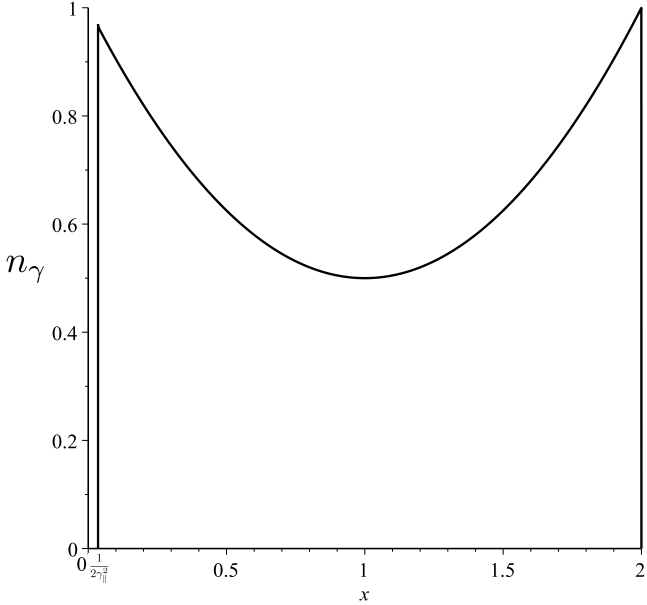}
\caption{The frequency distribution of the photons. The symmetrical form of the profile corresponds to the symmetry of the angular distribution of the radiation (see Fig. \ref{fig:CR_angle}).}
\label{fig:counts}
\end{figure}

3) In the case of large $\gamma_\parallel \gg 1$ the width of the beam is much smaller than the telescope aperture, and
the RDR radiation coming from all the beam is registered, with the spectrum  (\ref{psi12}).
Denoting $x=\frac{\omega}{\gamma_\parallel \omega_B}$, the frequency distribution in the registered signal (\ref{psi12})
may be written as

\begin{equation}
\label{ppar4}
W(\omega)\sim \tilde W(x)=x(1-x+\frac{x^2}{2}), \quad \frac{1}{2 \gamma_\parallel^2}<x<2.
\end{equation}
Here $\tilde W_{max}=2$ at $x_{max}=2$, the median emissivity $W=1$ is reached at $x_{med}\approx 1.52$, the 50\% and 90\% of emission
is concentrated inside $1.53<x<2$, and $0.63<x<2$ respectively, when

\begin{equation}
\label{ppar5}
\frac{W}{W_{tot}}=\frac{\int_{x_{50,90}}^2 x(1-x+\frac{x^2}{2})dx}{\int_0^{2} x(1-x+\frac{x^2}{2})dx}=0.5,\,\,0.9
\end{equation}
respectively. The observed average frequency of the RDR therefore, corresponds to $x\approx 1.5$. The effective width of the line
is about $\delta\omega \approx 0.25 \omega_{max}$. The form of the line on the photon counts figure $n_\gamma(\omega)$ follows the law

\begin{equation}
\label{ppar6}
n_\gamma(x) = 1-x+\frac{x^2}{2},\quad {\rm at}\quad \frac{1}{2 \gamma_\parallel^2}<x<2,
\end{equation}
with a smooth slope at lower frequencies and abrupt brake at $x=2$, $\omega=2\gamma_\parallel \omega_B$. Stress, that this form of the line is
present for monoenergetic electron beam with $\gamma_\parallel$ equal for all electrons. This type of widening takes place only in
the RDR mechanism, and the form of the magneto-dipole line may be used for distinguishing between the cyclotron and RDR mechanisms.
Note that in presence of other widening mechanisms the line may become even wider, but should preserve characteristic feature of this
mechanism. The form of the line $\tilde W(x)$ and the frequency distribution of the photons are shown on Figs \ref{fig:line} and \ref{fig:counts}.

\subsection{Polarization properties}

Linear $W_l(\omega)$ and circular $W_c(\omega)$ parts of the total radiation, using (\ref{psi2b}), (\ref{psi2c}), (\ref{psi12}), may be expressed in the following way

\begin{equation}
\label{ppar7}
W_l(\omega)=W(\omega)\rho_l^2(\omega)=\frac{e^{2} v_{\perp 0}^{2}\omega_B}{2c^{3}\gamma_\parallel^2}
x^3\frac{(1-\frac{x}{2})^2}{1-x+\frac{x^2}{2}},
\end{equation}

\begin{equation}
\label{ppar8}
W_c(\omega)=W(\omega)\rho_l^2(\omega)=\frac{e^{2} v_{\perp 0}^{2}\omega_B}{2c^{3}\gamma_\parallel^2}
x\frac{(1-x)^2}{1-x+\frac{x^2}{2}}.
\end{equation}
Here $x=\frac{\omega}{\gamma_\parallel \omega_B}$, $0<x<2$. It is easy to show that $W_l(\omega)+W_c(\omega)=W(\omega)$ from (\ref{psi12}).
The linear polarization of the total radiation in the beam $W_l$ is obtained by integration over the frequency, it gives

\begin{equation}
\label{ppar9}
W_l=\int_0^2 W_l(\omega)dx= \frac{e^{2} v_{\perp 0}^{2}\omega_B}{2c^{3}\gamma_\parallel^2}\int_0^2 x^3\frac{(1-\frac{x}{2})^2}{1-x+\frac{x^2}{2}}dx
=
\end{equation}
$$
=(\pi-\frac{8}{3})\frac{e^{2} v_{\perp 0}^{2}\omega_B}{2c^{3}\gamma_\parallel^2}.$$
The linearly polarized part of the radiation in the beam $RDR_l$ is defined as

\begin{equation}
\label{ppar10}
RDR_l=\frac{W_l}{W_{tot}}=\frac{3}{4}\int_0^2 x^3\frac{(1-\frac{x}{2})^2}{1-x+\frac{x^2}{2}}dx=\frac{3}{4}(\pi-\frac{8}{3})
\approx
\end{equation}
$$\approx \frac{3}{4}\cdot 0.475\approx 0.356.$$
 The circularly polarized part of the beam radiation has different sign in the wavelength intervals $\frac{\omega_B}{2\gamma_\parallel}<\omega<\gamma_\parallel\omega_B$ ($0<x<1$), and $\omega_B\gamma_\parallel<\omega<2\gamma_\parallel\omega_B$ ($1<x<2$).
The circularly polarized part of the radiation at lower frequencies $RDR_{c1}$ ($0<x<1$) is determined as

$$RDR_{cl}=\frac{W_{c1}}{W_{tot1}}
$$
$$
=\left[\int_0^1 x\frac{(1-x)^2}{1-x+\frac{x^2}{2}}dx\right]/\left[\int_0^1 x(1-x+\frac{x^2}{2})\right]$$
\begin{equation}
\label{ppar11}
=(\ln 2-\frac{\pi}{2}+1)/\frac{7}{24}=\frac{0.122}{0.292}\approx 0.418.
\end{equation}
The circularly polarized part of the radiation at higher frequencies $RDR_{c2}$ ($1<x<2$) has an opposite sign of polarization, and is determined as

$$RDR_{c2}=\frac{W_{c2}}{W_{tot2}}
$$
$$
=\left[\int_1^2 x\frac{(1-x)^2}{1-x+\frac{x^2}{2}}dx\right]/\left[\int_1^2 x(1-x+\frac{x^2}{2})\right]$$
\begin{equation}
\label{ppar12}
=(3-\ln 2-\frac{\pi}{2})/\frac{25}{24}=\frac{0.736}{1.042}\approx 0.707.
\end{equation}

\section{Discussion}

The analysis of two model of the magneto dipole mechanism of the line formation had shown important features, by measuring of which
the cyclotron model may be distinguished from the RDR model with strongly anisotropic electron distribution. The angular distribution
in these models is very different, quasi-isotropic in the cyclotron model, and strongly anisotropic in RDR. Nevertheless, in observing
these, usually rather weak lines, it is hardly possible to obtain a definite answer about the angular distribution, due to many contaminating factors.
Different form of the line, and different observed light curves, predicted by theory for these lines could help more,  when measured, but for
weak lines it seems also very difficult.
  The most distinct difference between the cyclotron and RDR mechanisms of the line formation may be seen in their polarization features.
The measurements of the hard  X-ray polarization are discussed for more than 40 years, but still there is no space mission for
these measurements. The linear X-ray polarization which is measured should be close to zero for the cyclotron radiation from the hot
magnetic pole, while the radiation produced in RDR model should have about 35\% of the linear polarization in the line, what is
sufficiently well distinguished difference, which could be taken into account in the construction of the hard X-ray polarimeter.
The first source with the detected magnetic dipole line Her X-1 is still the best target for this investigation.

The line emitted by non-relativistic electrons in the magnetic field has the same cyclotron frequency. Its harmonics are
 highly suppressed at $kT\ll m_e c^2$, what is expected in the accretion disk and in the accretion column. The observed line
 width originating from photons of the relativistic dipole radiation coming from different angles is forming one broad line,
 contrary to separate harmonics in the cyclotron model. The second harmonic in the case of relativistic dipole radiation should
 form another line of almost the same width at double energy. There is still not clear wether  the second "cyclotron" harmonics
 is present in the X-ray spectrum of  Her X1 (Tr\"umper at al., 1978; Enoto et al., 2008; F\"urst et al., 2013).

The influence of the line broadening due to the distribution over the parallel electron momentum is considered by
Baushev and Bisnovatyi-Kogan (1999). According to (\ref{ppar3}), the broadening is small when the scattering momentum is non-relativistic.
With increasing of $\sigma$ over $m_e c$, the broadening increases, and the resulting line broadening is the combination of the angular
and scattering action. Nevertheless, it is very important that for any scattering in a  parallel direction the parameters of the
 polarization do not  change for any large $\gamma_{\parallel}$, which are considered in this model,
 and the criteria for the choice between models remains valid.

\section*{Acknowledgements}

This work was partially supported by the Russian Foundation for Basic Research Grant No. 14-02-00728 and the Russian Federation President Grant for Support of Leading Scientific Schools, Grant No. NSh-261.2014.2.
The work of GSBK was partially supported by the Russian Foundation for Basic Research Grant No. OFI-M 14-29-06045. The work of YaSL was partially supported by the Dynasty Foundation.









\bsp	
\label{lastpage}
\end{document}